\documentstyle[prl,epsf,aps]{revtex}

\newcommand \be {\begin{equation}}
\newcommand \bea {\begin{eqnarray}}
\newcommand \ee {\end{equation}}
\newcommand \eea {\end{eqnarray}}
 
 \newcommand \bi {\bibitem}

\newcommand \De {\Delta}

\newcommand \al {\alpha}

\newcommand \dd {{\mathrm{d}}}
\makeatletter
\newcommand\erfc{\mathop{\operator@font erfc}\nolimits}
\makeatother
\input{epsf}
\begin{document}
\twocolumn[\hsize\textwidth\columnwidth\hsize\csname@twocolumnfalse\endcsname
\draft       

\title{Aging in the linear harmonic oscillator}
\author{L. L. Bonilla$^{(*)}$, F. G. Padilla$^{(*)}$ and 
F. Ritort$^{(**)}$} 
\address{
(*) Departamento de Matem{\'a}ticas,\\
Universidad Carlos III, Butarque 15\\
Legan{\'e}s 28911, Madrid (Spain)\\
(**) Institute of Theoretical Physics\\ University of Amsterdam\\
Valckenierstraat 65\\ 1018 XE Amsterdam (The Netherlands).\\
E-Mail: bonilla@ing.uc3m.es,padilla@dulcinea.uc3m.es,ritort@phys.uva.nl}

\date{\today}
\maketitle

\begin{abstract}
The low temperature Monte Carlo dynamics of an ensemble of linear
 harmonic oscillators shows some entropic barriers related to the
difficulty of finding the directions in configurational space which
decrease the energy. This mechanism is enough to observe some typical 
non-equilibrium features of glassy systems like activated-type behavior
and aging in the correlation function and in the response function. 
Due to the absence of interactions the model only displays a 
one-step relaxation process.
\end{abstract} 

\vfill

\vfill

\twocolumn
\vskip.5pc]

\narrowtext
Slow relaxation processes are widespread in condensed matter
physics. These include magnetic relaxation in spin glasses, transport
processes in structural glasses, pinning effects in superconductors
among others. A large class of these systems show what is commonly
referred as aging, i.e. dependence of the response of the system on the
time in which it is perturbed. Aging effects \cite{sue} are a signature that
the system is far from thermal equilibrium and consequently the 
fluctuation-dissipation theorem is not valid \cite{REV1}. It has
been realized quite recently that aging is indeed a solution of the
off-equilibrium dynamics in some exactly solvable models \cite{CUKU,FrMe}. 
Aging appears if relaxation to the equilibrium is
slow due to the presence of energy barriers in a rugged free energy
landscape as well as in systems with entropy barriers with a quite
simple landscape \cite{BG1,MEBA}. In this last case, as the system relaxes
towards the equilibrium, the number of directions in phase
space where the system can move decreases progressively.
This means that the system needs more time to decorrelate 
or to forget the previous
configuration. This effect is usually encoded in the two time
correlation function where the $C(t,t')$ depends on both time 
indices \cite{Rieger}.

From previous considerations it is clear that aging can also
be present in extremely simple relaxing systems without any interaction,
the only condition being the progressive reduction of available
phase space where the energy decreases. This was an essential ingredient
in the Backgammon model recently proposed to explain  glassy
behavior in the absence of energy barriers \cite{BG1}. Here we consider
a simpler example and analyze the
Brownian oscillator. The Brownian oscillator is usually studied in
the Langevin approach. It is described in any textbook of stochastic
theory \cite{KAM}. It is possible to show that in this case there are 
no slow processes involved. In fact, the relaxation turns out to be
exponential as expected for the dynamics of a particle in a
single parabolic potential well. Here we consider the Monte Carlo approach
and choose a dynamics based on the Metropolis
algorithm \cite{METR}. This Monte Carlo approach was already 
studied in a disordered
model with long-range interactions which turns out to be non trivial,
at least in the zero temperature limit \cite{mSK}. The simplest case
of an harmonic oscillator is solvable and we analyze the dynamics here.

In \cite{mSK} we checked that, after
a suitable rescaling of time, the equilibrium Langevin and Monte Carlo 
dynamics are equivalent. Also, we showed how the Langevin dynamics can
be derived from the Monte Carlo dynamics in the limit of small
changes. Here, we will see that the same results are valid.
We will obtain the dynamical equations for the energy, the correlation
and the response function. We will also study the low temperature dynamics,
showing the similarities and differences with more realistic models
for glasses.

The harmonic oscillator has an energy,

\be
E=\frac{1}{2}Kx^2
\label{eq1}
\ee

\noindent
where $K$ is the Hooke constant and $x$ defines the position of the
harmonic particle. Let us consider an ensemble of $N$ independent linear
oscillators with total energy $E(\lbrace
x_i\rbrace)=(K/2)\sum_ix_i^2$. A change of $\{x_i\}$ is proposed
$\{x_i\to x'_i=x_i+r_i/\sqrt{N} ,\forall i\}$ where $\{r_i\}$ is
randomly chosen for each oscillator from a Gaussian distribution of zero
mean and finite variance equal to $\Delta^2$. The change is accepted
with probability $1$ if the energy decreases, i.e. if $\delta
E=E(\lbrace x'_i\rbrace)-E(\lbrace x_i\rbrace)$ is negative. Otherwise
the change is accepted with probability $\exp(-\beta\delta E)$ where
$\beta=\frac{1}{T}$ is the inverse of the temperature of the heat bath.

Let us sketch the main derivation of the dynamical quantities \cite{mSK}. We
first consider the probability that a given set of movements $\{x_i\to
x'_i=x_i+r_i/\sqrt{N}\}$ changes the energy in a quantity $\delta E$. This
probability is given by,

\newpage

\bea P(\delta E)=\int_{-\infty}^{\infty} \delta(\delta
E-K\sum_i(\frac{r_ix_i}{\sqrt{N}}+\frac{r_i^2}{2N})) \nonumber \\ \Bigl(
\prod_i \frac{\dd r_i}{\sqrt{2\pi\Delta^2}}
{-\frac{1}{2}}\exp(-\frac{r_i^2}{2\Delta^2})\Bigr )\label{eq2} \eea

For simplicity we have considered the case in which the mean position
$M=\frac{1}{N} \Sigma_{i=1}^{N} \langle x_i \rangle$ of the initial
condition is zero.  The average $\langle... \rangle$ is done over
different dynamical histories starting with the same initial condition
for the ensemble.  Using the integral representation for the delta
function in the thermodynamic limit $N\to\infty$ we obtain,

\be
P(\delta E)=(4\pi K E \Delta^2)^{-\frac{1}{2}}\exp\bigl(-\frac{(\delta
E-\frac{K\Delta^2}{2})^2}{4KE\Delta^2}\bigr )
\label{eq3}
\ee

Because the probability distribution $P(\delta E)$ only depends on the
energy itself the dynamics is then Markovian and simple to
solve. Obviously this result is solely due to the simplicity of the
model. According to the Metropolis dynamics the equation of evolution for
the energy is,

\be
\frac{\partial E}{\partial t}=\int_{-\infty}^0\dd x\,x P(x)+\int_0^{\infty}
\dd x\,x P(x) \exp(-\beta x)
\label{eq4}
\ee

\noindent 
which yields

\bea
\frac{\partial E}{\partial t}=\frac{a_c}{2}\Bigl ( \frac{1-4E\beta}{a_c
\beta} f(t)+ \erfc(\al)\Bigr)
\label{eq5}
\eea

\noindent
where $\al=(K\Delta^2/16E)^{\frac{1}{2}}$, $a_c=\frac{\Delta^2 K}{2}$ and

\bea
\erfc(x)=(2/\sqrt{\pi})\int_x^{\infty}\exp(-x^2)\dd x\\
f(t)=a_c\beta\,e^{-\beta a_C(1-2E(t)\beta)}\,
\erfc\bigl (\al(t)(4E(t)\beta-1)\bigr )
\label{eq11}
\eea

\noindent
It is easy to check that the only
stationary solution of this dynamical equation corresponds to the
equilibrium solution with $E=T/2$ (in agreement with the equipartition
theorem). The equation (\ref{eq5}) is already closed and yields the dynamical
evolution of the energy at all times.

Knowing the evolution of the energy we can calculate the acceptation
rate. This is defined by

\bea
A(t)=\int_{-\infty}^0\dd x P(x)+\int_0^{\infty}
\dd x P(x) e^{-\beta x}=
\nonumber
\\
\frac{1}{2} (\frac{f(t)}{a_c \beta} + \erfc(\alpha))
\label{eqac}
\eea

\noindent
In equilibrium, we have $E_{eq}=\frac{T}{2}$, and the  
acceptation rate becomes $A_{eq}=\erfc(\alpha_{eq})$. 

Similarly one can derive equations for the correlation $C(t,t')$ and response
function $G(t,t')$ (from now on, we will consider the first time $t'$ as the
smallest one, i.e. $t'<t$) defined by,

\bea
C(t,t')=\frac{1}{N}\langle \sum_i x_i(t')x_i(t)\rangle\\
G(t,t')=\Bigl (\frac{\delta M(t)}{\delta h(t')}\Bigr )_{h=0}~~~~~t'<t
\label{eq6}
\eea

\noindent
where $M(t)=(1/N)\sum_i\langle x_i(t)\rangle$ is the average position of
the ensemble of oscillators. The response function is computed with the
energy $E=(K/2)\sum_i x_i^2 -h\sum_i x_i$ starting from an initial
condition $M=0$ at zero field and taking finally the limit $h\to 0$. In
the case that $M(t=0)\ne 0$ the computation is more involved and the
equation of motion for the response function $G(t,t')$ involves also the
energy response function $G_E(t,t')=\Bigl (\frac{\delta E(t)}{\delta
h(t')}\Bigr )_{h=0}$. It is not difficult to obtain the equations for both
correlation and response functions. Proceeding in a similar way as for
the case of the energy we get,

\bea
\frac{\partial C(t,t')}{\partial t}=-f(t)C(t,t')\label{eq10a}\\
\frac{\partial G(t,t')}{\partial t}=-f(t)(G(t,t')-
\frac{1}{K}\delta(t-t'))\label{eq10b}
\eea

\noindent
where $\al(t)$  and $f(t)$ were previously defined. 
The difference in the equations for the $C$ and $G$ concern only
the initial condition. Note that again
the Markovian properties of the dynamics are manifest because the time
derivative of the $C$ (or $G$) depends solely on the $C(t,t')$ (or
$G(t,t')$) itself and the initial conditions $C(t',t')=2 E(t')/K $,
$lim_{t \to (t')^-}G(t,t')=0$.

We can easily integrate the equations for the correlation and response
functions. We obtain that they depend on the energy (through the
function $f(t)$) at all previous times,

\bea
C(t,t')=\frac{2E(t')}{K}\exp(-\int_{t'}^t\,f(x)\dd x)\label{eqc1}\\
G(t,t')=\frac{f(t')}{K}\exp(-\int_{t'}^t\,f(x)\dd x)\Theta(t-t')\label{eqg1}
\eea

With these exact results we can also calculate the
fluctuation-dissipation parameter,

\begin{eqnarray}
X(t,t')=\frac{\frac{\partial C(t,t')}{\partial t'}}{TG(t,t')}=1-2E'/T
+\nonumber\\
\frac{\erfc(\alpha') \exp(a_c \beta (1-2 E'\beta))}
{\erfc(\alpha' (4 E' \beta -1))}\label{eqX}
\end{eqnarray}

\noindent
where $\alpha'=\alpha(t')$ and $E'=E(t')$. Note that $X(t,t')$ only
depends on the smallest time $t'$.
  
It is straightforward to check that in thermal equilibrium both
correlation and response only depend on the difference of times, the
fluctuation-dissipation theorem $TG(t)=-\dot{C}(t)$ is satisfied and the
$X(t,t')=1$. This is a general consequence of the detailed balance
property inherent to the Metropolis algorithm.  With an appropriate
rescaling of time we find also that the equilibrium form of correlation
and response function are the same as in the Langevin case. Equations
(\ref{eq5}) and (\ref{eq10a},\ref{eq10b}) are much different that the
corresponding ones in the Langevin dynamics. In particular, the response
function at equal times is $1$ in the Langevin dynamics while in the
Monte Carlo case it is different from 1, even in thermal equilibrium.

The Langevin case is a
limit of the Monte Carlo dynamics. This result has been demonstrated in
the framework of the SK spherical spin-glass model \cite{mSK} and also
applies here. When the size of the typical movement $\Delta$ goes to zero
equations (\ref{eq5},\ref{eq10a},\ref{eq10b}) become,

\bea
\frac{\partial E}{\partial t}=\frac{K\De^2}{2}(1-2E\beta)\label{eq11a}\\
\frac{\partial C(t,t')}{\partial t}=-\frac{K\beta\De^2}{2}C(t,t')
\label{eq11b}\\
\frac{\partial G(t,t')}{\partial t}=-\frac{K\beta\De^2}{2}(G(t,t')-\frac{1}{K}\delta(t-t'))
\label{eq11c}
\eea

\noindent
with initial conditions $C(t,t') = K E(t')/2$ and $lim_{t \to (t')^-}
G(t',t')=0$. These are the same equations as in the Langevin dynamics 
with uncorrelated white noise with variance $2T$ if the time is rescaled
by the quantity $\De^2\beta/2$. This means that both dynamics are
essentially equivalent in case the rescaling factor $\De^2\beta/2$ is
finite. The interesting case corresponds to the low temperature limit 
$\beta\to\infty$ for $\Delta$ fixed. In this case a new
relaxational dynamics driven by a low acceptance rate is
found. Linearization of eq.(\ref{eq5}) around the equilibrium solution
yields a relaxation time which diverges at low temperatures like
$\tau\simeq \beta^{\frac{1}{2}}\exp(K\De^2\beta/8)$. This implies a
divergence of the relaxation time of an activated type. 

\begin{figure}
\begin{center}
\leavevmode
\epsfysize=230pt{\epsffile{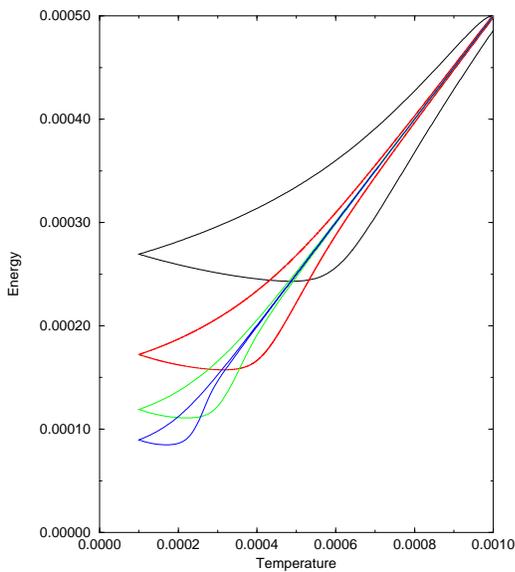}}
\end{center}
  \protect\caption[1]{ Cooling experiment. Values of the energy when we
decrease and increase the temperature of the system at different ratios.
From top to bottom, ratios 0.1,0.01,0.001 and 0.0001.
    \protect\label{FIG2}
  }
\end{figure}

At very low temperatures the
harmonic oscillator wants to relax to a configuration of very small
entropy (indeed,  because the oscillator is classical, the entropy
diverges like $log(T)$ at low temperatures). 
In this situation the oscillator spends the
major part of time looking for the ground state configuration. In
some sense, the dynamics generates itself entropy barriers in a single
potential well. This means that if we
perform a cooling experiment, decreasing and increasing the temperature
at a fixed rate, we expect that the system fails to relax to the
equilibrium energy (see figure \ref{FIG2}). This is a typical feature of glassy
systems.

Let us consider the evolution of the energy at zero temperature. In this
case, only those changes $\{\delta x_i\}$ which decrease the energy are
accepted. Close to the equilibrium point $x=0$ the system will relax
very slowly, mainly because the largest part of the movements are
rejected. The relaxation of the energy at zero temperature is given by,

\be
\frac{\partial E}{\partial
t}=-(\frac{KE\De^2}{\pi})^{\frac{1}{2}}\exp(-\al^2)\,+\,\frac{K\De^2}{4}
\erfc(\al)
\label{eq12}
\ee

To obtain the long time behavior we expand the error function in the 
limit $\al\to\infty$,

\be
\frac{\partial E}{\partial t}=-\bigl (\frac{64}{\pi K\De^2}\bigr
)^{\frac{1}{2}}E^{\frac{3}{2}}\exp(-\frac{K\De^2}{16 E})
\label{eq13}
\ee

We should note that previous equation is extremely similar to that
derived in the Monte Carlo dynamics of the Sherrington-Kirkpatrick
spherical model in the adiabatic approximation at zero temperature. In
terms of the parameter $\al$ (defined after eq.(\ref{eq5})) the
equation (\ref{eq13}) can be written in the simple form,

\be
\frac{\partial \al}{\partial t}=\frac{\exp(-\al^2)}{\sqrt{\pi}}~~~~.
\label{eq130}
\ee

For large times, the parameter $\al$ grows logarithmically in time,

\be \al(t)\simeq \bigl
(\log(\frac{2t}{\sqrt{\pi}})+\frac{1}{2}\log(\log(\frac{2t}{\sqrt{\pi}})\bigr)^{\frac{1}{2}}
\label{eqal}
\ee

\noindent the acceptation rate eq.(\ref{eqac}) decays like

\be
A(t)\simeq \frac{1}{4t\log(\frac{2t}{\sqrt{\pi}})}
\label{eqac2}\ee

\noindent plus subdominant logarithmic corrections. The energy also decays 
logarithmically in time

\be E(t) \simeq \frac{K \Delta^2}{16}
\frac{1}{\log(\frac{2t}{\sqrt{\pi}})+ \frac{1}{2}
\log(\log(\frac{2t}{\sqrt{\pi}}))}
\label{eq131}
\ee
\noindent 
Similarly the correlation function satisfies the equation,

\be
\frac{\partial C(t,t')}{\partial t}= -\bigl( \frac{K\De^2}{4\pi}\bigr
)^{\frac{1}{2}}  \frac{C(t,t')}{\sqrt{E(t)}}\exp(-\frac{K\De^2}{16
E(t)})
\label{eq14}
\ee

The same equation is fulfilled for the response function.
Using the asymptotic differential equations for the energy (\ref{eq13}),
the correlation (\ref{eq14}) and the response function,
we can show that the correlation function for long times displays 
a solution of the type $C(t,t')\simeq \frac{2 E(t')}{K} \frac{g(t)}{g(t')}$
and $G(t,t')\simeq \frac{2 \alpha(t')}{\sqrt{\pi} K} g(t)$, 
being $g(t)=\exp(-\al^2)$. Using the asymptotic 
expression for the energy (\ref{eq131}) we get,
\bea
C_{norm}(t,t')=\frac{C(t,t')}{2 E(t')/K}\simeq 
\frac{t'}{t} \Bigl
(\frac{\log(\frac{2t'}{\sqrt{\pi}})}{\log(\frac{2t}{\sqrt{\pi}})}\Bigr )^{\frac{1}{2}}
\label{eq15}\\
G(t,t')\simeq \frac{1}{Kt}\Bigl (\frac{\log(\frac{2t'}{\sqrt{\pi}})+
\frac{1}{2}\log(\log(\frac{2t'}{\sqrt{\pi}}))}{\log(\frac{2t}{\sqrt{\pi}})}
\Bigr )^{\frac{1}{2}}\label{eqr}
\eea

\noindent 

\begin{figure}
\begin{center}
\leavevmode
\epsfysize=230pt{\epsffile{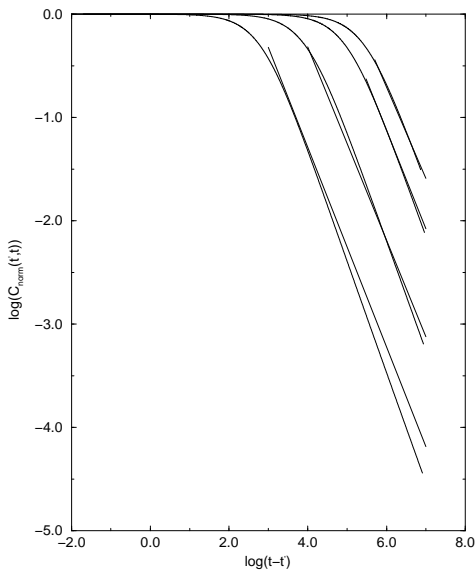}}
\end{center}
  \protect\caption[2]{Correlation function for different waiting times.
From top to bottom, $t'=3\cdot10^5$,$1\cdot10^5$,$1\cdot10^4$,$1\cdot 10^3$.
The short lines show the calculated asymptotic behavior eq(\ref{eq15})
    \protect\label{FIG1}
  }
\end{figure}

This approximation is valid in the asymptotic limit of large values of
 $t'$. The normalized correlation function shows aging behavior with a
 simple scaling form $t'/t$ plus some logarithmic
 corrections. Apparently the response function (\ref{eqr}) does not show
 aging because it does not depend on a ratio of functions depending on
 $t$ and $t'$. But this is an artifact of the normalization factor
 necessary to make the response function to take a finite value at
 equal times. In fact, the leading behavior of the response function
 decays to zero for large values of $t$ and an appropriate normalization
 of the response function at equal times is necessary (in the same way
 as has been done for the correlation function). Note that for Langevin
 dynamics the normalization of the response function is not necessary
 since the $G(t,t')$ at equal times already takes a finite value by
 definition (e.g., $\lim_{t'\rightarrow (t)^-}G(t,t')=1$). The
 normalized response function takes the simpler form,

\be G_{norm}(t,t')=\frac{G(t,t')}{G(t'+0,t')}\simeq \frac{t'}{t}\Bigl
(\frac{\log(\frac{2t'}{\sqrt{\pi}})}{\log(\frac{2t}{\sqrt{\pi}})}\Bigr
)^{\frac{1}{2}}
\label{eqgnorm}
\ee

which displays aging with the same leading behavior as the
normalized correlation function. We can obtain information about the
dynamics (and in particular, about the response function) from the
remanent magnetization \cite{vinc}. In the present model the
magnetization corresponds to the average position of the ensemble of
oscillators (defined after (\ref{eq6})). The main equations
(\ref{eq5}),(\ref{eq10a}) and (\ref{eq10b}) have been derived in the
absence of external field and starting with zero initial
magnetization. In this case it is natural to compute the zero-field
cooled magnetization. In this procedure the system is suddenly cooled
down to a given temperature and after a waiting time $t_w$ a small
step field $h(t')=h\Theta (t'-t_w)$ is applied. If the value of $h$ is
small enough then we are in the linear response regime. The
magnetization starts to grow according to the relation,

\bea
M_{ZFC}=\int_{-\infty}^{t}G(t,t')h(t')dt'=\nonumber\\
=h\int_{t_w}^{t}G(t,t')dt'=hI(t_w,t)
\label{eqmzfc}
\eea

The quantity $I(t_w,t)=\int_{t_w}^{t}G(t,t')dt'$ defines the integrated
response function. From the exact expression obtained for the response
function (\ref{eqg1}) we get, for the integrated response function,

\be
I(t_w,t)=\frac{1}{K}\bigl( 1-\exp(-\int_{t_w}^tf(t')dt')\bigr )
\label{I}
\ee

In the large time limit $t\to\infty$ the zero-field cooled magnetization
converges to its equilibrium value, the field-cooled magnetization
$M_{FC}$. It is easy to check, from (\ref{eqmzfc}), (\ref{I}) that
$M_{FC}$ is given by $M_{FC}=\frac{h}{K}$, ie, the equilibrium linear
magnetic susceptibility $\chi_0=M_{FC}/h=\frac{1}{K}$ is independent of
the temperature. 

Another interesting quantity to be calculated is the anomaly in the
response function \cite{meza}, defined as,

\bea
{\bar \chi} \equiv \lim_{t \rightarrow \infty} 
\int_0^t dt' G(t,t')-\int_0^{\infty} G_{eq}(\tau) d\tau
\\
=\lim_{t \rightarrow \infty} I(0,t)-\chi_0\\
=-\frac{1}{k} e^{-\int_0^t f(t') dt'}
\label{anomaly}\\
G_{eq}(\tau)=\lim_{t_w \rightarrow \infty} G(t_w,t_w+\tau)
\eea  

For a finite value of $\beta$, the system decays to equilibrium
in a finite time and for long times the integral $\int_0^t dt' f(t')$
behaves as $f_{eq}t$. This implies that the magnetization relaxes
exponentially to zero, showing no aging for large values of $t_w$ and
the 'anomaly' relaxes exponentially to zero too. The behavior of the
anomaly and the zero-field cooled magnetization is more interesting at
zero temperature. In that case it can be shown that the leading behavior
of the anomaly decays algebraically (as $1/Kt$) to zero. Using the
asymptotic behavior of the energy eq.(\ref{eq131}), it is easy to
check that the zero-field cooled magnetization goes like,

\be
\frac{M_{ZFC}}{M_{FC}}\simeq 1-\frac{t_w}{t}\Bigl
(\frac{\log(\frac{2t_w}{\sqrt{\pi}})}{\log(\frac{2t}{\sqrt{\pi}})}\Bigr
)^{\frac{1}{2}}
\label{zfc}
\ee

Using the linear response relation $M_{ZFC}+M_{TRM}=M_{FC}$ where
$M_{TRM}$ is the thermo-remanent magnetization obtained by quenching the
system in an (small) applied field and removing it at $t_w$, we get

\be
\frac{M_{TRM}}{M_{FC}}\simeq \frac{t_w}{t}\Bigl
(\frac{\log(\frac{2t_w}{\sqrt{\pi}})}{\log(\frac{2t}{\sqrt{\pi}})}\Bigr
)^{\frac{1}{2}}~~~~~.\label{trm}
\ee

Both $M_{ZFC}$ and $M_{TRM}$ show aging with the leading
$t/t_w$ scaling behavior. In figure 3 we show the thermo-remanent
magnetization for the oscillator model for different values of $t_w$.

\begin{figure}
\begin{center}
\leavevmode
\epsfysize=230pt{\epsffile{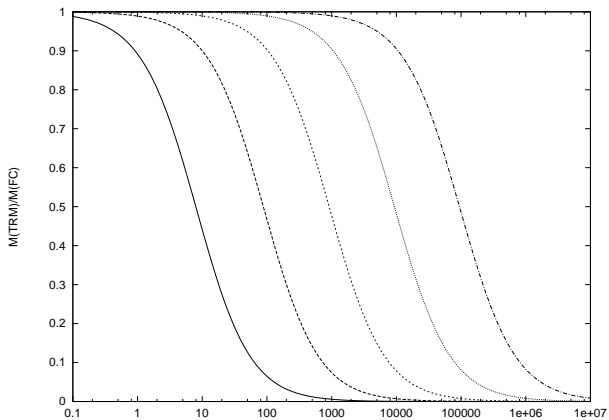}}
\end{center}
  \protect\caption[2]{Thermo-remanent magnetization for different
waiting times obtained from eq.(\ref{trm}). From left to right,
$t_w=10$,$10^2$,$10^3$,$10^4$,$10^5$.  \protect\label{FIG3} }
\end{figure}

It has been suggested that the $X(t,t')$ could be interpreted as an
effective temperature \cite{BG2,CKP}. If we define $T_f(t')=X(t,t')/T$
then, from eq. (\ref{eqX}), the fluctuation-dissipation theorem is
obeyed with the effective temperature $T_f(t')$. While this is a formal
relation it would be interesting if the effective temperature derived in
this way had some deep physical meaning. On the other hand, a
well-founded physical interpretation of the violation of the
fluctuation-dissipation relation, to our knowledge, does not exist. Note
that it is possible to define different fluctuation-dissipation ratios
(for instance, $X(t,t')=-TG/\frac{\partial C}{\partial t}$) all giving
$X(t,t')=1$ in equilibrium. The definition here adopted is the
conventional one which allows to obtain a closed expression for the
integrated response function in case the function $X$ is solely function
of the correlation $C(t,t')$ \cite{CUKU,FrMe}. On
general physical grounds one would expect an effective temperature
larger than the temperature of the bath. To raise the temperature should
contribute (by the equipartition theorem) those degrees of freedom
which, during the process of relaxation towards the equilibrium, still
are not frozen. For the simple model considered here such an
interpretation seems to work.  From equation (\ref{eqX}) it can be shown
that the effective temperature for a system relaxing at zero temperature
is given by the relation $T_f(t')\to 2E(t')$. Consequently the effective
temperature and the dynamical energy in the off-equilibrium regime are
related by the thermodynamic relation suggesting that some kind of
adiabatic theorem holds for this simple system in the long time
limit. One can then ask if the whole time dependent probability
distribution $p_t(\lbrace x_i\rbrace)$ in the long-time limit is of the
Boltzmann type but dependent on an effective temperature $T_f(t)$,
i.e. $p_t(\lbrace x_i\rbrace)\sim \exp(-E(\lbrace
x_i\rbrace)/T_f(t))$. It is easy to check that such a result is not
possible \cite{FRRI} and equipartitioning is valid only for some finite
moments of the probability distribution (for instance the second moment,
i.e. the energy).

In conclusion, we have studied the Monte Carlo dynamics of an ensemble
of linear harmonic oscillators. The extreme simplicity of this model
makes it exactly solvable without loosing the interesting features of
the non-equilibrium dynamics driven by entropic barriers. In this way,
we are able to gather quite a lot of information and derive all relevant
dynamical quantities with reasonable analytical effort. 
 
We find a very slow relaxation near zero temperature, driven by a low
acceptance rate, similar to that found in the Backgammon model
\cite{BG1}, models of adsorption \cite{AS} as well as models for
compaction of dry granular media \cite{CLHN}. In these cases, the origin
of the slow relaxation is the existence of entropic barriers, although
they are set up by different mechanisms. Note that the notion of
entropic barrier or entropic trap is quite similar to the concept of
{\em effective volume} in free volume theories. In our case this
manifests as a inverse logarithmic law decay of the energy
eq.(\ref{eq131}) while in compaction of granular media this decay is
found for the density of compaction. The model has also in common with
models for glasses aging in the correlation function for long times. The
correlation function $C(t,t')$ presents a $\frac{t'}{t}$ behavior with
some logarithmic corrections (with $t'$ the smallest time in the
correlation function).  It is interesting to note that this corrections
appear also in the Backgammon model \cite{GL} (and presumably also in
adsorption models \cite{AS} and models for compaction of dry granular
media \cite{CLHN}) but do not appear in models with Langevin dynamics
\cite{CUDE,llp}).  We have found also aging in the magnetization (the
integrated response function). This behavior appears associated to the
algebraical decay of the 'anomaly' as $t$ goes to infinity.  Due to the
zero value of the anomaly we expect a finite value of the overlap between two
replicas \cite{BBM} in the large $t$ limit if putted in the same
configuration at $t_w$ (cloning procedure). This expectation stems from
the simplicity of the landscape in this model (a single parabolic
well).  Consequently, this model falls into the first dynamical
category (class I) proposed in \cite{BBM}. However this model shares
some features of the $p$-spin model (with $p > 2$, belonging to class
II) like aging in the integrated response function.
Furthermore, this simple
model lacks a fast process decaying to a plateau and also a two time
dependence of the fluctuation-dissipation parameter, which only depends
on the smallest time. This is probably due to the absence of a
cooperative behavior.

{\bf Acknowledgments}. We are grateful to S. Franz and
Th. M. Nieuwenhuizen for a careful reading of the manuscript. The work
by L.L.B. and F.G.P.  has been supported by the DGES of Spain under
grant PB95-0296. The work by F.R has been supported by FOM under
contract FOM-67596 (The Netherlands).

\hspace{-2cm}

\vfill

\end{document}